\documentclass[aps,pre,reprint,superscriptaddress]{revtex4-1}
\usepackage[utf8]{inputenc}
\usepackage{graphicx}
\usepackage{placeins}
\usepackage{floatflt}
\usepackage{natbib}
\usepackage{epsfig,epsf,rotating}
\usepackage{amsmath}
\usepackage{amsopn}
\usepackage{amssymb}
\usepackage{bm}
\usepackage[normalem]{ulem}
\graphicspath{ {chap4Figures/} }

\DeclareMathOperator{\Tr}{Tr}

\begin{document}
	\bibliographystyle{apsrev4-1}
	\title{Combined  molecular dynamics and quantum trajectories simulation of laser-driven, collisional systems}
	\author{G.M. Gorman}
	\affiliation{Rice University, Department of Physics and Astronomy, Houston, Texas, USA}
	\author{T.K. Langin}
	\affiliation{Rice University, Department of Physics and Astronomy, Houston, Texas, USA}
	\author{M.K. Warrens}
	\affiliation{Rice University, Department of Physics and Astronomy, Houston, Texas, USA}
	\author{D. Vrinceanu}
	\affiliation{Texas Southern University, Department of Physics, Houston, Texas, USA}
	\author{T.C. Killian}
	\email[]{killian@rice.edu}
	\affiliation{Rice University, Department of Physics and Astronomy, Houston, Texas, USA}
	
	\date{\today}
	
	\begin{abstract}
		We introduce a combined molecular dynamics (MD) and quantum trajectories (QT) code to simulate the effects of near-resonant optical fields on state-vector evolution and particle motion in a collisional system. In contrast to collisionless systems, in which the quantum dynamics of multi-level, laser-driven particles with spontaneous emission can be described with the optical Bloch equations (OBEs), particle velocities in sufficiently collisional systems change on timescales comparable to those of the laser-induced, quantum-state dynamics. These transient velocity changes can cause the time-averaged velocity dependence of the quantum state to differ from the OBE solution.
We use this multiscale code to describe laser-cooling in a strontium ultracold neutral plasma. Important phenomena described by the simulation include suppression of electromagnetically induced transparencies through rapid velocity changing collisions and thermalization between cooled and un-cooled directions for anisotropic laser cooling.
	\end{abstract}
	
	
	\maketitle

\section{Introduction}
\label{sec:Introduction}

Laser-generated forces on atoms, ions, and molecules, such as in laser cooling \cite{mva2003}, arise from coupling of external momenta and internal quantum states of the particles of interest. In most cases, optical forces can be calculated using the velocity-dependent, steady-state solutions to the optical Bloch equations (OBEs) for internal-state quantum dynamics \cite{lukin}. In a highly collisional system, however, the particle velocities and associated Doppler shifts can change significantly on the timescale required for the internal quantum states to reach steady state.  These rapid  velocity changes can cause the time-averaged  quantum state, and thus the calculated optical forces, to differ from  the steady-state OBE solution. One such collisional system is ions in an ultracold neutral plasma (UNP) \cite{kkr1999,kpr2007,lro2017}. Laser cooling of ions in an UNP  was recently demonstrated in \cite{lgk2019}.

UNPs are typically created by photo-exciting laser-cooled atoms just above the ionization threshold. The temperature of resulting electrons ($T_e \sim 1-1000$\,K) is set  by the detuning of the ionization laser above threshold. The ion temperature is set by equilibration dynamics after plasma formation \cite{kpr2007,sck2004} and is typically below 1 K. Ions in an UNP are strongly coupled, meaning the average Coulomb interaction energy between neighboring ions is larger than the thermal kinetic energy, and standard kinetic descriptions of the evolution of the velocity distribution become invalid \cite{gms2002,ppr2004PRA}. For an accurate description of ion dynamics in UNPs, direct molecular dynamics (MD) simulations must be used (\textit{e.g.} \cite{lsk2016}), which evolve the motional dynamics of individual particles under the influence of inter-particle interactions.

Here, we introduce a computational code that couples a MD simulation with a  quantum trajectories (QT) description of internal-state dynamics \cite{dcm1992,car1992}. To describe a collisional laser-driven system, the QT algorithm evolves the internal quantum state and calculates the optical forces for each individual ion based on its velocity and internal state, and the inter-ion forces are derived from the MD algorithm. The velocities and positions of the ions evolve under the influence of both forces.  This code is used to investigate laser cooling of ions in an UNP. The MDQT code is multiscale in the sense that it couples the fast, internal quantum dynamics to the classical motion of the ensemble of particles. A similar computational tool for evolving the quantum state in a collisional system was described in \cite{dbr2009}, but in that work optical forces were not taken into account in particle kinematics.

\begin{figure*}
	\includegraphics[width=1\textwidth]{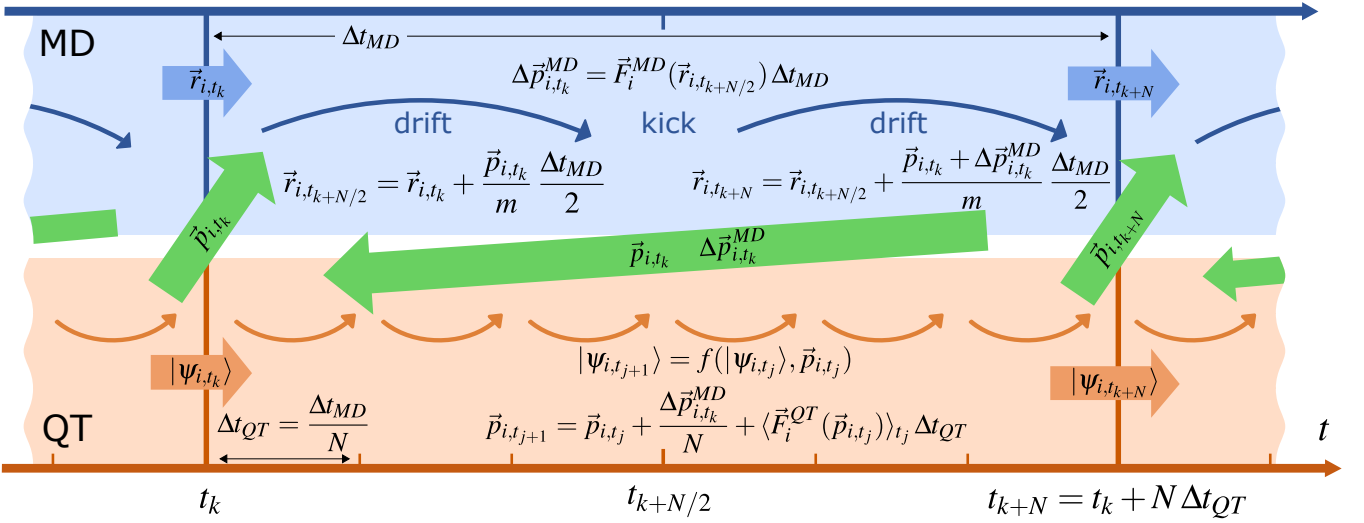}
	\caption{
		Combined simulation for an ensemble of particles under the influence of a classical, position-dependent force ($\vec{F}^{MD}$) and a   momentum-dependent force ($\vec{F}^{QT}$) that depends upon the internal quantum state. The state vectors $(|\psi\rangle{})$ are evolved  within a quantum trajectories (QT) algorithm (orange) with a time step $\Delta t_{QT}$, and the positions $(\vec{r})$ are  evolved within a  position-Verlet (leapfrog) molecular dynamics (MD) algorithm (blue), which treats the classical force with a time step $\Delta t_{MD}=N \Delta t_{QT}$, where $N$ is an integer. The momenta  $(\vec{p})$, however, are evolved in both the quantum and classical realms. The MD and QT algorithms share a common time axis, and the subscript $i$ is the particle-label index. A combined MDQT time step begins with a single MD time step, after which the initial momenta and classical-force momentum kicks $(\Delta \vec{p}^{MD})$ for the current time step are passed to the QT code. This initiates evolution of state vectors and momenta during the corresponding series of $N$ QT time steps. $\Delta \vec{p}^{\,MD}$ is spread evenly across the QT time steps to reduce error in calculations of $\vec{F}^{QT}$. Following completion of the $N$ QT time steps, the final momenta are passed  to the MD algorithm to initiate the next combined MDQT step. Equations for the MD algorithm are indicated. The function  that evolves the particle state vectors ($f$) is described in Sec.\,\ref{sec:QT}.
	}
	\label{fig:combMdAndQt}
\end{figure*}

The paper is structured as follows. Section \ref{sec:combMDandQTgeneral} provides a general overview of the architecture developed for simulating the dynamics of an ensemble of particles under the influence of classical, position-dependent forces and  momentum- and quantum-state-dependent forces.
In Sec.~\ref{sec:MDSims}  we describe the application of MD to a system of ions interacting through a  screened Coulomb interaction, which is appropriate to describe the ion dynamics of interest in an UNP. In Sec.~\ref{sec:QT} we introduce the QT simulation and specific details for describing laser-driven Sr$^{+}$ ions. In Sec.\ \ref{sec:darkstatesl}, we discuss dark-state formation in laser-driven Sr$^{+}$ ions, comparing numerical solutions of the OBEs and of the MDQT code. The latter shows the suppression of dark-states in the collisional environment of an UNP.
In Sec.~\ref{sec:SimResults} we show the results from a MDQT simulation of laser-cooling of the UNP and compare results with experimental data \cite{lgk2019}. We conclude in Sec.~\ref{sec:SimSummary}.

\section{Combining Classical and Quantum Simulations}
\label{sec:combMDandQTgeneral}
Figure \ref{fig:combMdAndQt} provides a  schematic of the general architecture for simulating the dynamics of an ensemble of particles under the influence of classical, position-dependent forces ($\vec{F}^{MD}$) and momentum- and quantum-state-vector-dependent forces ($\vec{F}^{QT}$).
The classical forces are treated with a MD code that uses a position-Verlet (leapfrog) integrator with time step $\Delta t_{MD}$, while $\vec{F}^{QT}$ and the state-vector evolution are treated with a QT code using a time step of $\Delta t_{QT}$. For the physical system of interest here, the quantum state dynamics are typically faster than the kinematics, and we assume that the $\Delta t_{MD}/ \Delta t_{QT}\equiv N\ge 1$, where $N$ is an integer.  Sections \ref{sec:MDSims} and \ref{sec:QT}, respectively describe the specific MD and QT algorithms used and how simulation parameters relate to the UNP experiment.

Particle positions $(\vec{r})$, momenta $(\vec{p})$, and state vectors $(|\psi\rangle{})$ are the fundamental quantities evolved within the combined MDQT code.
Forces calculated in the MD portion of the algorithm depend on the positions of all particles, as is typically the case for MD simulations.
The state vectors are evolved  within the QT code and the positions are evolved within the MD code. The momenta, however, are shared between both the quantum and classical realms.

The MD and QT algorithms share a common time axis, but they are not run simultaneously.
A combined MDQT time step begins with a MD time step, which evolves particle positions and determines the classical-force momentum kicks $(\Delta \vec{p}^{\,MD})$ according to the position-Verlet (leapfrog) equations shown in Fig.\,\ref{fig:combMdAndQt}. While the updated positions are stored for the next MD time step, the initial momenta and classical momentum kicks are  passed to the QT code. This initiates a series of $N$ QT time steps corresponding to the same time interval as the MD step, during which the state vectors and particle momenta are evolved.

The
initial state vector at the start of each QT time step is taken from the output of the previous step, and this is used to calculate the expectation value of $\vec{F}^{QT}$.
The state vector evolution  and quantum-force calculation are described in Sec.\ \ref{sec:QT}. Each particle's momentum is changed during each QT time step by $\Delta \vec{p}^{\,MD}/N$ plus the impulse resulting from the quantum force calculated for that QT step. In this way, the classical-force momentum kick is spread evenly across the QT steps, reducing error in  calculations of $\vec{F}^{QT}$.
 Following completion of the $N$ QT time steps, the evolved momenta are passed back to the MD code and the process repeats.

The architecture described here is multiscale in the sense that it links numerical simulations of classical and quantum dynamics that typically occur on different time scales. It is well suited for a computationally expensive classical MD component, which is often the case for a many-body classical force (see Sec.\,\ref{sec:MDSims}).

%

\section{Molecular Dynamics Simulation}
\label{sec:MDSims}

In MD simulations, one numerically solves Hamilton's equations of motion
for an $N$-body system of pair-wise interacting particles with potential typically of the form $V({r}_{ij})$, for distance ${r}_{ij}$ between particles $i$ and $j$.
These techniques \cite{haile} were first applied to hard-spheres \cite{awa1959} and liquids interacting through a Lennard-Jones potential \cite{rah1964,ver1967}, before being applied in plasmas \cite{hmp1975}. We refer to the forces obtained with this classical calculation as $\vec{F}^{MD}$.


The ion dynamics of interest here can be described with a Yukawa one-component plasma (YOCP) model \cite{fha1994,mur2004} in which particles interact through a screened, repulsive $1/r$ potential (Eq.~\ref{YukawaInteraction})
\begin{equation}
  V(r_{ij})=\frac{e^{2}}{4\pi\epsilon_{0}r_{ij}}\exp\left(-\frac{r_{ij}}{\lambda_{D}}\right).
 \label{YukawaInteraction}
\end{equation}
  Electrons serve as a neutralizing and screening background, introducing the Debye screening length $\lambda_{D}=\sqrt{k_{B}T_{e}\epsilon_{0}/(ne^{2})}$, where $n$ is the density and $T_{e}$ is the electron temperature. This approach neglects electron-ion thermalization \cite{msk2015} and three-body recombination \cite{kpr2007}, which are good approximations for our conditions. These effects could in principle be added at various levels of approximation.

The YOCP model is commonly used to describe  plasmas,  especially
under conditions of strong coupling \cite{hfd1997,fha1994,mur2006} such as for  white dwarf stars \cite{sva1969}, the
cores of Jovian planets \cite{ste1980,rdr2006}, plasmas produced during inertial confinement fusion \cite{lin1995}, dusty plasmas consisting of highly charged dust particles \cite{smm2014,miv2009}, and ions in UNPs \cite{sck2004,kkr1999}.

The MD algorithm used here evolves a YOCP of typically $N=3500$ particles in a cube of volume $L^3$ with uniform density  and periodic boundary conditions using the minimum image convention (MIC)~\cite{fsm2000} and a position-Verlet (leapfrog) integrator \cite{hea1981} of Hamilton's equations of motion. The natural time step for the MD simulation is $\Delta t_{MD}=0.0017/\omega_{pi}$, for ion plasma oscillation frequency $\omega_{pi}=\sqrt{ne^{2}/\epsilon_{0}m_{i}}$ with $n$ and $m_i$ the ion density and mass respectively.
The initial conditions for the particles are random positions and zero kinetic energy, simulating the initial conditions for typical UNP experiments.
More details on the MD simulation can be found in \cite{lsk2016,LanginPHD}

When using periodic boundary conditions with the MIC, the forces from image charges other than the nearest neighbor are ignored. Due to the $\exp[-\kappa r]$ term in (Eq.~\ref{YukawaInteraction}), the potential for a YOCP system depends strongly on the plasma screening parameter $\kappa=a_{ws}/\lambda_{D}$,
where $a_{ws}=(3/4\pi n)^{1/3}$ is the Wigner-Seitz radius. Thus, it is important that the system size be large enough such that the force exerted by image charges other than the nearest image is negligible.  In general, the condition for MIC validity can be written as $L\kappa \gg 1$.  In~\cite{fha1994MIC}, convergence in the observed melting point of a Yukawa solid was demonstrated for a number of particles $N_{conv}\approx 435/\kappa^{3}$. We perform simulations for $\kappa\approx 0.5$, for which $N_{conv}\approx 3500$. For typical conditions of simulations used here, energy is conserved at better than the 10$^{-4}$ level.

In a combined MDQT simulation, the traditional MD algorithm is modified such that the momenta are updated by the quantum-state-vector-dependent forces in between each MD time step, as described in Sec.\ \ref{sec:combMDandQTgeneral}.

\section{Quantum Trajectories}
\label{sec:QT}
\subsection{Introduction}
\label{sec:QTIntro}
\begin{figure*}
	\includegraphics[width=.8\textwidth]{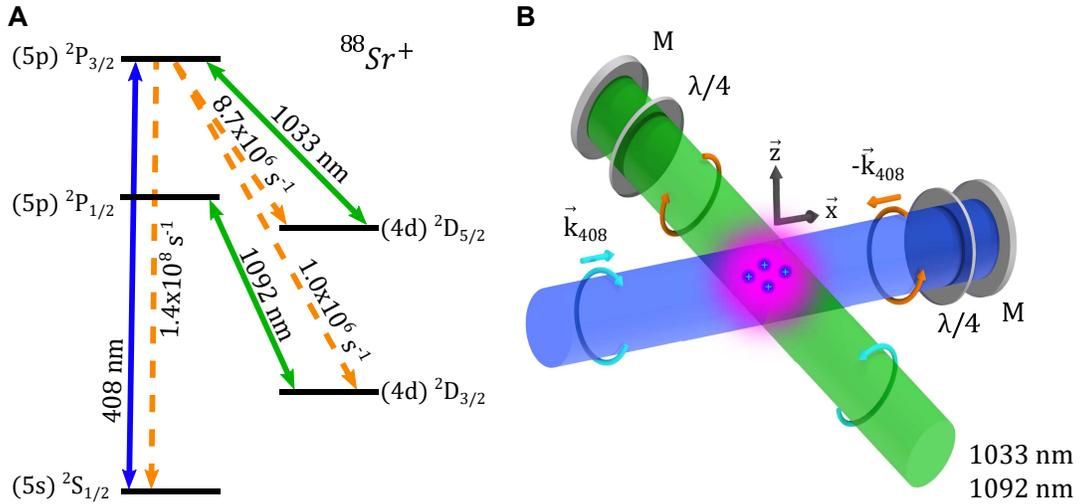}
	\caption{
(A)	Sr$^+$ level diagram including wavelengths and decay rates for transitions relevant to laser cooling.
(B)	Experimental schematic. 
Cooling (408\,nm) and repumping (1033\,nm, 1092\, nm) lasers are applied in counter-propagating configurations with indicated polarizations. Propagation directions for cooling beams are indicated.  (M: mirror, and $\lambda/4$: quarter-wave plate). Adapted from \cite{lgk2019}.
}
	\label{fig:expApparatus}
\end{figure*}

The quantum trajectories method  \cite{dcm1992,car1992,lukin} utilizes an equivalence between the master equation, which describes the time evolution of a single-particle, pure-state density matrix $\rho=|\psi\rangle\langle\psi|$ in an open quantum system, and the evolution of a wavefunction $|\psi\rangle$ under an equivalent \emph{non-Hermitian} Hamiltonian.  At any time step, $|\psi\rangle$ can also jump via spontaneous emission to  ground states $|\phi\rangle$ with a probability proportional to that of occupying an excited state.

The master equation for the evolution of a pure quantum state, in its most general form, can be written
\begin{eqnarray}
  \frac{d\rho}{dt} &=& \frac{1}{i\hbar}\left[H^{QT},\rho\right]-\sum_{k}\frac{\gamma_{k}}{2}\left(c_{k}^{\dagger}c_{k}\rho+\rho c_{k}^{\dagger}c_{k}-2c_{k}\rho c_{k}^{\dagger}\right)\nonumber \\
   &=& \frac{1}{i\hbar}\left[H^{QT}_{eff},\rho\right]+\sum_{k}\gamma_{k}c_{k}\rho\,c_{k}^{\dagger}
   \label{eq:MasterEqGeneral}
\end{eqnarray}
 where $c_{k}$ are quantum jump operators with associated rates $\gamma_{k}$ (e.g. $k$ indexes each possible decay path, so $c_{k}\equiv|\beta\rangle\langle \alpha|$ if the initial and final states for decay path k are $|\alpha\rangle$ and $|\beta\rangle$, respectively), and $H^{QT}$ is the system Hamiltonian, which is independent of coupling to the reservoir/vacuum and $H^{QT}$ describes some process that can stimulate transitions between internal states, such as stimulated emission and absorption due to near-resonant laser fields.
The first term on the RHS of the second line of Eq.\ \ref{eq:MasterEqGeneral}
corresponds to the evolution of a pure state $|\psi\rangle$ under the \emph{non-Hermitian} Hamiltonian $H^{QT}_{eff}=H^{QT}-i\hbar\sum_{k}\frac{\gamma_{k}}{2}c_{k}^{\dagger}c_{k}$. 
The second term on the RHS  of the second line of Eq.\ \ref{eq:MasterEqGeneral} handles quantum jumps that change $|\psi\rangle$ into another, properly normalized state 
$|\phi_{k}\rangle=\sqrt{\gamma_{k}\Delta t_{QT}/\Delta P_{k}}c_{k}|\psi\rangle$,
which are caused by the coupling to the external environment that results in, for example, spontaneous emission. Here,
\begin{equation}
\Delta P_{k}(t)=\Delta t_{QT}\gamma_{k}\langle\psi(t)|c_{k}^{\dagger}c_{k}|\psi(t)\rangle.
\label{eq:dpk}
\end{equation}
In the situation of interest here, $H^{QT}$ and thus $H^{QT}_{eff}$ contain interactions arising from the optical fields involved in laser cooling ions in the plasma (App.\ \ref{app:QTdetails}).

The evolution of the state vector during one time step from time $t$ to  $t+\Delta t_{QT}$ is numerically calculated as follows. During each time step  the state vector either jumps into one of the $|\phi_{k}\rangle$ states
with probability $\Delta P_{k}(t)$
or the state vector evolves for time $\Delta t_{QT}$ according to $H^{QT}_{eff}$.
The probability that the wavefunction ``jumps'' during the time step is given by
\begin{equation}
\Delta P\equiv
\sum_{k}\Delta P_{k}(t).
\label{eq:dp}
\end{equation}

For numerical efficiency, we evolve the state vector in the case of no jump using a 4$^{th}$-order Runge-Kutta method. This is implemented by approximating  the evolution of the state vector from $t$ to $t+dt$ with
\begin{equation}
|\psi(t+dt)\rangle = \frac{1+H^{QT}_{eff}(t)dt/i\hbar}{\sqrt{1-\Delta P(t)}}|\psi(t)\rangle
\label{eq:psiEv}
\end{equation}
where any time-varying terms in the RHS of Eq.\ \ref{eq:psiEv}  are evaluated at $t$.

The internal state dynamics and the time-varying classical momentum of the particle, $\vec{p}(t)$, are coupled. The momentum determines the Doppler shifts for light fields, which are taken into account in $H^{QT}$.
In a time step in which there is a quantum jump, the momentum changes due to the discrete recoil momentum kick associated with photon emission accompanying the quantum jump transition. In a time step without a jump, the momentum evolves under the influence of the optical force, which can be calculated at any time as
\begin{equation}
\langle \vec{F}^{QT}(t)\rangle=\Tr \left({\rho}\frac{d{\vec{p}}}{dt}\right) = \Tr \left({\rho}\frac{\left[{\vec{p}},H^{QT}_{eff}\right]}{i\hbar}\right).
\label{eq:ForceProfileFromTrace}
\end{equation}
This treats particle momentum classically, and we find it sufficient to evolve the momentum during a no-jump time step $\Delta t_{QT}$ with the Euler method. If there is an additional force, not associated with the optical fields,  its action on particle momentum during time step $\Delta t_{QT}$ (Eq.\,\ref{dpx,elec}) can be included by adding the resulting momentum change to the impulse from a quantum jump or the optical force. The classical momentum kicks ($\Delta \vec{p}^{MD}$) are treated in this fashion as described in Sec.\ \ref{sec:combMDandQTgeneral}.

To describe an ensemble of particles, as required to describe laser cooling of ions in an UNP, we evolve and  track the state vector and momentum for each particle, $|\psi_{i}(t)\rangle$ and $\vec{p}_i(t)$. As needed, for each particle, we form the density matrix, $\rho_i(t)$ from $|\psi_{i}(t)\rangle$. For the situation of interest here, the QT evolution is not sensitive to positions of the particles.  More details can be found in \cite{LanginPHD}.

The QT  algorithm was validated \cite{LanginPHD} by reproducing analytic  results for simple two- and three-level systems for single-particle phenomena such as Rabi oscillations and ensemble properties such as the cooling rate and cooling limit for Doppler-cooling.

\subsection{Applying Quantum Trajectories to the Laser Cooling of $^{88}$Sr Ions}
\begin{figure*}
	\centering
	\includegraphics[width=.8\textwidth]{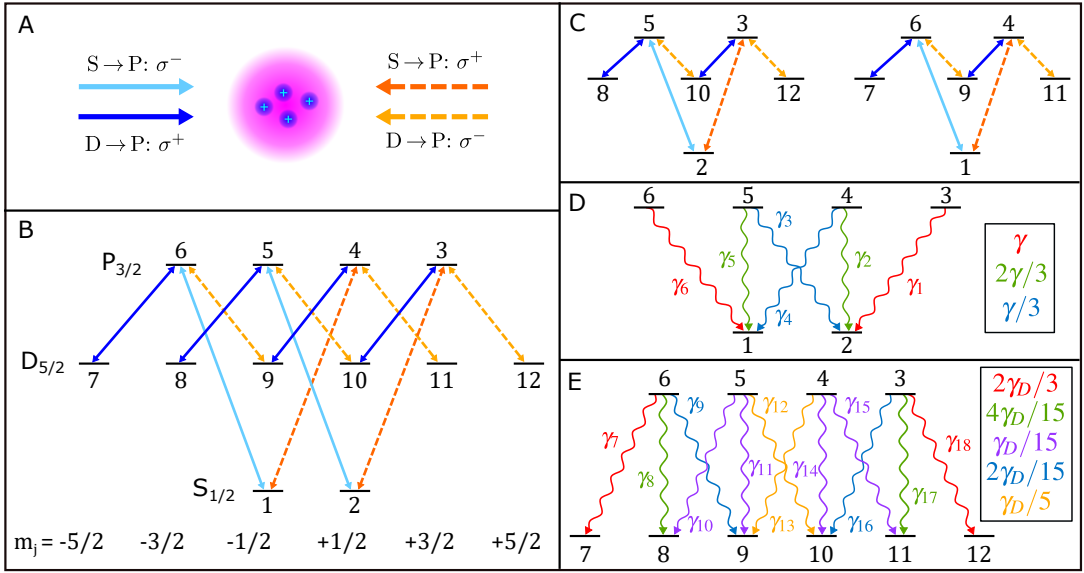}
	\caption[Diagram Illustrating Full Sr$^{+}$ Level Diagram With Zeeman Sublevels]{(\textbf{A}): Counter-propagating, cross-polarized laser configuration considered in the QT simulation.  Dashes indicate leftward propagating lasers, which are Doppler-shifted by $+kv_{x}$.  (\textbf{B}): Corresponding level diagram including full Zeeman substructure with the indicated laser-couplings. (\textbf{C}): If the states are coupled only by circularly polarized light, the 12-level system can be separated into two 6-level subsystems, making it easier to recognize dark states induced by near-resonant two-photon coupling between combinations of S and D sublevels (see Sec.~\ref{subsec:OBESol}).  (\textbf{D}): Decay channels from P to S, with decay rates indicated ($\gamma=1.41\times10^{8}$\,s$^{-1}$).  (\textbf{E}): Decay channels from P to D, with decay rates indicated ($\gamma_{D}=8.7\times10^{6}$\,s$^{-1}$). Numerical labels of the decay channels in (D) and (E) correspond to the labels for the jump operators in Eq. \ref{eq:MasterEqGeneral}.}
	\label{fig:FullSrLevelDiagram}
\end{figure*}

To  model laser cooling of ions in an UNP, $H^{QT}$ takes the form appropriate for the level structure of  $^{88}$Sr ions and the laser configuration of ref.\ \cite{lgk2019}. Figure \ref{fig:expApparatus} shows the Sr$^+$ level diagram and experimental schematic for one-dimensional laser cooling
on the D2 line (${^{2}S_{1/2}}\rightarrow {^{2}P_{3/2}}$) at 408\,nm in Sr$^+$ \cite{lgk2019}. This transition is not closed, and repumping lasers must be added to remove population from the $^{2}D_{j}$ states. For simplicity, in the simulation the repump lasers are oriented along the axis of the cooling lasers, which we define as the $x$-axis, as shown in
in Fig.~\ref{fig:FullSrLevelDiagram}A. We define $\delta$ and $\delta_D$ as the detunings of the 408-nm and 1033-nm lasers from resonance for an ion at rest.


Figure~\ref{fig:FullSrLevelDiagram}B shows all the levels treated in the QT calculation
including Zeeman substructure.
We ignore decay into the $^{2}D_{3/2}$ state and the corresponding 1092-nm repump laser.  This is justified due to the small branching ratio (1:151) into this state (compared to 1:17 for the $^{2}D_{5/2}$ state) and because ions that fall into the $^{2}D_{3/2}$ state are repumped via the $^{2}$P$_{1/2}$ level, not the $^{2}$P$_{3/2}$ level.  This is sufficient for relatively short simulations, over which the $^{2}D_{3/2}$ level remains largely unpopulated.  For longer simulations, it is straightforward to graft a rate equation approach onto the QT code to treat population dynamics involving the $^{2}D_{3/2}$ state. The resulting effective Hamiltonian and additional details of the QT implementation are given in App.\ \ref{app:QTdetails}.

For simulating UNP laser cooling, the expectation value of the laser-induced force is calculated. In the particular geometry considered here, forces are only
along one dimension, $\langle \vec{F}^{QT}_i\rangle=\langle F^{QT}_{xi}\rangle\hat{x}$, where
\begin{equation}
\langle F^{QT}_{xi}\rangle=\Tr \left({\rho}_i\frac{d{p}_{xi}}{dt}\right) = \Tr \left({\rho}_i\frac{\left[{p}_{xi},H^{QT}_{eff}\right]}{i\hbar}\right)
\label{eq:ForceProfileFromTrace}
\end{equation}
is the $x$-component of the optical force on  ion $i$ at a particular time and ${p}_{xi}$ is the momentum along $x$.  The explicit form of $\langle F^{QT}_{xi}\rangle$ is given in
Eq.\ \ref{eq:Forcei} \cite{supplementary}, but it depends on ${p}_{xi}$ through the Doppler shift of the laser frequencies. $\langle F_{xi}^{QT}\rangle$ is independent of particle positions because the Coulomb interactions shift all internal states equally. Thus collisional broadening of the optical transitions is negligible.

In every time step of the QT simulation, the state vector and momentum  of each particle are evolved as described in Sec.\ \ref{sec:QTIntro}.
%
%
%
If a quantum jump occurs for an ion, its $x$-momentum receives a recoil kick of
$\pm \hbar k$ or $\pm\hbar k_{D}$, where $k$ and $k_D$ are the wave vectors corresponding to the photon emitted during
a jump from a $P$ state to an $S$ state or a $D$ state respectively \footnote{Collisions rapidly isotropize the ion velocity distribution in the UNP, so no significant error is introduced with this poor approximation of a dipole radiation pattern.}.
When conducting a combined MDQT simulation, the momentum kicks calculated from the MD component of the code are included in the momentum evolution as described in Sec.\,\ref{sec:combMDandQTgeneral}.

The QT algorithmic details and values for physical parameters used in simulations presented here are given in
Sec.\ \ref{subsec:executeQTalgorithm}. The natural timescale for the QT simulation is set by the lifetime of the ${^2P_{3/2}}$ state, $\gamma^{-1}$, and the timestep is chosen as $\Delta t_{QT}=0.01\gamma^{-1}$.



The natural QT time step is typically smaller than the natural MD time step for typical UNP densities of $10^{16}$\,m$^{-3}$ or less ($ \frac{\Delta t_{MD}}{\Delta t_{QT}}=\frac{0.0017\omega_{pi}^{-1}}{0.01\gamma^{-1}}=\frac{17}{\sqrt{n}}$ where $n$ is in units of $10^{14}$\,m$^{-3}$). Thus, the fundamental time step for a combined MDQT simulation is taken as  $\Delta t_{QT}$. To account for numerical mismatch, the MD timestep is taken as $N=floor(\Delta t_{MD}/\Delta t_{QT})$ times $\Delta t_{QT}$.
For densities greater than $10^{16}$\,m$^{-3}$, which are rare in the Sr$^{+}$ UNP system, $0.0017\omega_{pi}^{-1}<0.01\gamma^{-1}$, in which case we reset $\Delta t_{QT}=\Delta t_{MD}=0.0017\omega_{pi}^{-1}$.
The simulation code used to produce all data within this report is made available via a GitHub repository \cite{mdqtCode}.

\section{Dark States for Laser-cooled Sr$^{+}$ Ions}
\label{sec:darkstatesl}
Dark states are eigenstates of the ion-light coupled Hamiltonian ($H^{QT}_{eff}$ without the decay terms) comprised of superpositions of only ${^2S_{1/2}}$ and ${^2D_{5/2}}$ states \cite{fim2005,dcm1992}.  A Sr$^{+}$ ion in a dark state does not scatter light, so population of these states may limit laser-cooling efficacy.

\subsection{Optical Bloch Equations (OBEs): Dark States for a Single Laser-cooled Sr$^{+}$ Ion}
\label{subsec:OBESol}

We first calculate internal-state populations and optical forces for a single ion using the optical Bloch equations (OBEs).
In a highly collisional ensemble of particles, like an UNP, rapid velocity changes may modify the time averaged populations and forces, but the OBEs provide important intuition and illustrate the effect of dark states.

Solving the OBEs amounts to solving the master equation for the evolution of the open quantum system (Eq.\,\ref{eq:MasterEqGeneral}), which we solve numerically assuming that at $t=0$ the population is all in the ground state.  After the steady state is reached, the optical force profile $F_{OBE}(v)=Tr({\rho} \left[p_x,H^{QT}(v)\right]/i\hbar\:)$, where we explictly indicate that $H^{QT}(v)$ depends on  the particle $x$-velocity $p_x/m \equiv v$ due to the Doppler shift, and populations of different internal states are determined. As mentioned previously, the force is only in the $x$ direction for the laser configuration considered here.

When ${^2S_{1/2}}\rightarrow {^2P_{3/2}}$ and ${^2D_{5/2}}\rightarrow{^2P_{3/2}}$ transitions are driven by $\sigma^{+}$ and $\sigma^{-}$ lasers, which we use for our simulations, the 12-level Sr$^{+}$ diagram separates into two subsystems of 6 levels each (See Fig.~\ref{fig:FullSrLevelDiagram}C).  The eigensolutions of the corresponding 6-level matrices are too complicated to include here.  Nevertheless, intuition can be gained by examining the subsystems.  Dark states typically exist when two states are coupled by a resonant two photon transition.  For example, a dark state is expected when the detunings of the photons coupling states $|2\rangle$ and $|3\rangle$ and states $|12\rangle$ and $|3\rangle$ cancel each other out, which occurs when $\delta-vk=\delta_{D}-vk_{D}$.  Similarly, the condition for the two photon coupling from $|2\rangle$ to $|8\rangle$ to be resonant is $\delta+vk=\delta_{D}+vk_{D}$.  There can also be dark states comprised solely of ${^2D_{5/2}}$ states, which are  all resonantly coupled for $v=0$. Dark states are thus expected at:

\begin{itemize}
	\item$v=0$ 
	\item $v=\pm (\delta-\delta_{D})/(k-k_{D})$
	\item $v=\pm (\delta-\delta_{D})/(k+k_{D})$ 
\end{itemize}

\begin{figure*}[t]
	\centering
	\includegraphics[scale=.65]{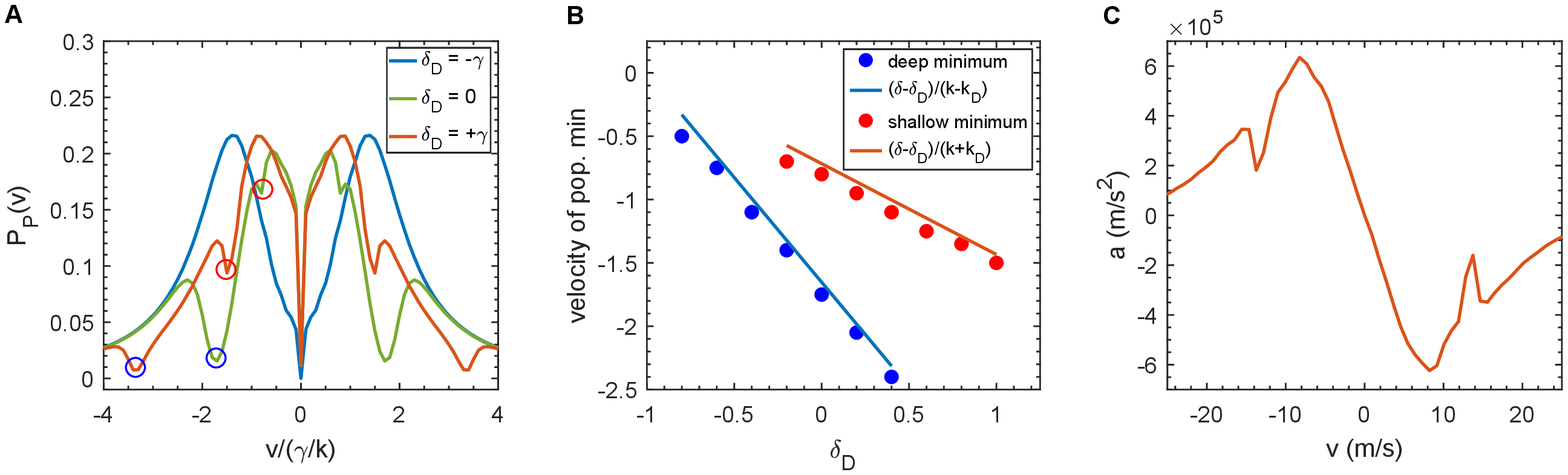}
	\caption[Observation of Dark States in OBE Solutions]{(\textbf{A}): ${^2P_{3/2}}$ state population, $P_p(v)$, as a function of ion velocity $v$ from steady state OBE solutions for various $\delta_{D}$ ($\delta = -\gamma$, $\Omega^{0}=\gamma$, and $\Omega^{0}_{D}=\gamma$ for all plots).  Minima in the ${^2P_{3/2}}$ state populations, circled in red and blue, correspond to velocity-dependent dark states. (Only the $v<0$ states are circled.) (\textbf{B}): Location of local minima in ${^2P_{3/2}}$ state populations vs $\delta_{D}$.  Locations closely match the velocities at which the two-photon transitions between ${^2S_{1/2}}$ and ${^2D_{5/2}}$ states are resonant.  (\textbf{C}): Acceleration profile $a(v)=F_{OBE}(v)/m$ obtained from steady-state OBE solution for $\delta=-\gamma$, $\delta_{D}=+\gamma$, $\Omega^{0}=\gamma$, and $\Omega^{0}_{D}=\gamma$.  Within the range defined by the capture velocity ($|v|<v_{c}=|\delta|/k=9$\,m/s), $a\propto -v$.}
	\label{fig:OBEPExc}
\end{figure*}

A  minimum in $P_{p}(v)$, the  population in the ${^2P_{3/2}}$ level, corresponds to significant population of a dark state.  In steady-state, the velocities at which this occurs depend on the laser detunings (Fig.~\ref{fig:OBEPExc}) and agree with the expected locations (Fig.~\ref{fig:OBEPExc}B).  The  acceleration due to laser forces ($a(v)=F_{OBE}(v)/m$) is  plotted in Fig.~\ref{fig:OBEPExc}C for $\delta=-\gamma$, $\delta_{D}=+\gamma$, $\Omega^{0}=\gamma$, and $\Omega^{0}_{D}=\gamma$.  In the region defined by $|v|\le 9$\,m/s, we find $a\propto-v$, as required for laser cooling. But at velocities for which there is significant dark-state population, the  laser acceleration displays minima, which can potentially reduce cooling effectiveness.

\begin{figure*}[t]
	\centering
	\includegraphics[scale=.7]{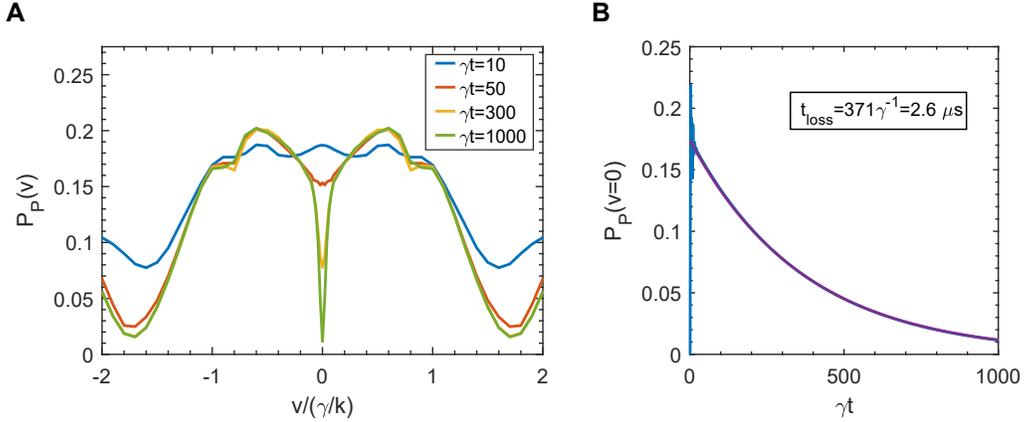}
	\caption[ $v=0$ Dark State Population Development Over Time]{(\textbf{A}): Time dependence of OBE solutions for $\delta=-\gamma$, $\delta_{D} = 0$, $\Omega^{0}=\gamma$, and $\Omega^{0}_{D}=\gamma$. The total time of the simulation in units of $\gamma^{-1}$ is indicated in the legend. The populations of $v=0$ dark states, comprised solely of ${^2D_{5/2}}$ state sublevels, develop quite slowly.  (\textbf{B}): ${^2P_{3/2}}$ state population at $v=0$ vs time. An exponential fit to the decay of population shows the timescale for the decay of ${^2P_{3/2}}$  population and development of corresponding dark states is  $\sim 2.6\mu$s. This is on the order of the timescale for velocity changing collisions ($\sim4\omega_{pi}^{-1}$) for a density of $1\times10^{14}$\,m$^{-3}$, and thus we may expect this state to be collisionally suppressed.
}
	\label{fig:DarkStateTimeDevelopment}
\end{figure*}

It is also worth considering how $P_{p}(v)$ depends on time.   Figure~\ref{fig:DarkStateTimeDevelopment}A shows $P_{p}(v)$ after the propagation of the OBEs for various lengths of time.  The different dark states develop at different rates, with the $v=0$ dark state taking the longest to develop.  $P_{p}(v=0)$ rises within a short time $\sim\gamma^{-1}$ and then decays exponentially with time (and thus the populations of the $v=0$ dark states grow) on a timescale of $t_{dark}\sim 370\gamma^{-1}=2.6\mu$s for typical laser cooling parameters (Fig.~\ref{fig:DarkStateTimeDevelopment}B). As we will now show (Sec.~\ref{sec:CollisionalDarkSupression}), this long timescale leads to collisional suppression of the dark state population, as ions are collisionally removed from near $v=0$ before they are optically pumped into the dark state.

\subsection{MDQT Simulation: Collisional Suppression of Dark States in a Laser-Cooled UNP}
\label{sec:CollisionalDarkSupression}

The combined MDQT code is useful for investigating the effects of collisions on the population of dark states during laser cooling.
In the absence of collisions, the dark state at $v=0$ (Fig.~\ref{fig:DarkStateTimeDevelopment}) is particularly slow to develop, with $t_{dark}\sim 2.6$\,$\mu$s, and also very narrow, with a velocity `full-width half-max' (FWHM) of $\delta v=0.6$\,m/s.  The velocity change in a time $dt$ due to collisions is given by $dv\sim \omega_{coll}v_{T}dt$, where $\omega_{coll}$ is the `velocity changing collision' (VCC) rate and is proportional to $\omega_{pi}$.  In \cite{bck2012}, $\omega_{coll}$ was measured to be $\sim 0.2\omega_{pi}$ for $\Gamma\sim 3$, as is the case in UNPs after equilibration.  The $v=0$ feature in $P_{p}(v)$ should be suppressed if $\omega_{coll}v_{T}t_{dark}>\delta v$, since the ion’s velocity changes by more than $\delta v$ within the 2.6 $\mu$s timeframe that it has to remain within $\delta v$ of zero velocity in order for it to relax into the dark state. Substituting in $\omega_{coll}=0.2\omega_{pi}$,  this is the case for $n\ge 10^{11}$\,m$^{-3}$.

\begin{figure*}[t]
	\centering
	\includegraphics[scale=.65]{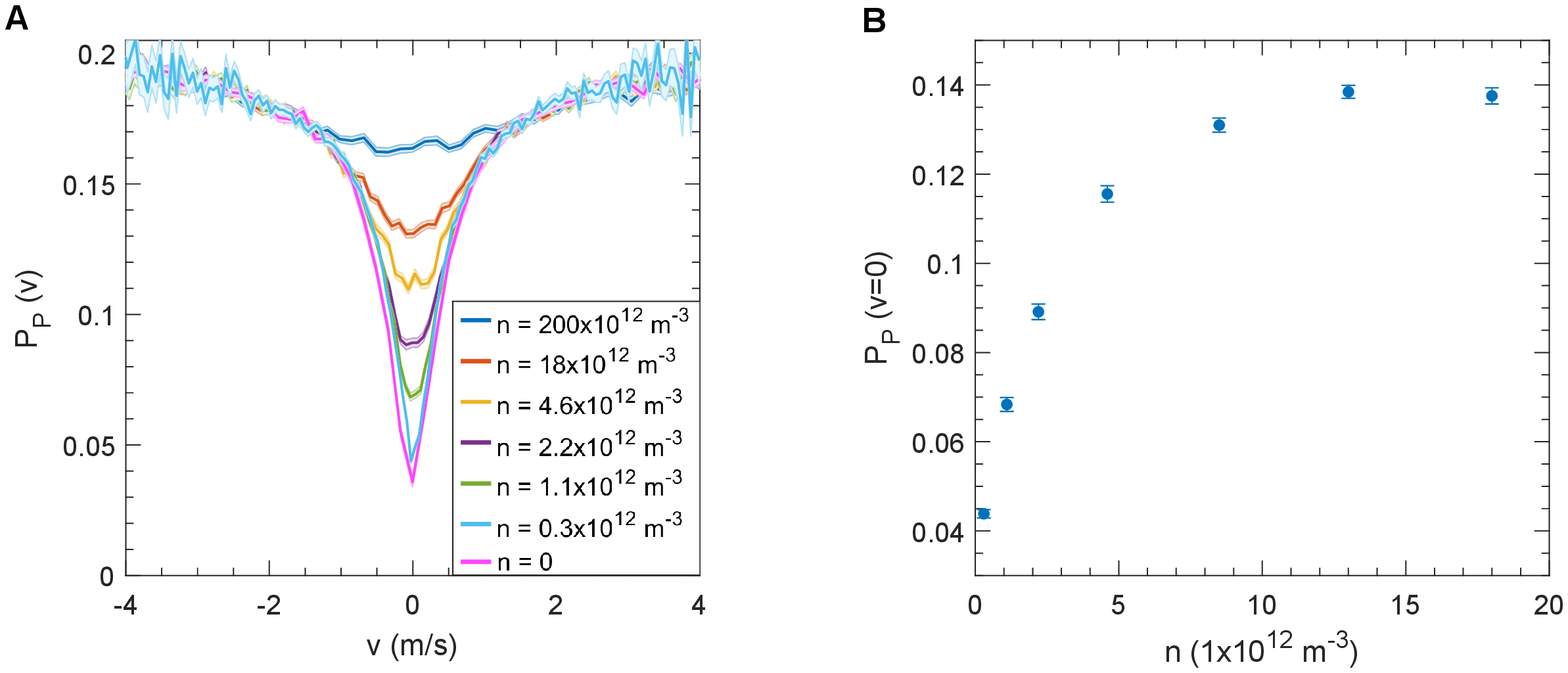}
	\caption[Collisional Supression of $v=0$ Dark State.]{(\textbf{A}):
	Predictions from the MDQT code for population in the ${^2P_{3/2}}$ state versus $x$-velocity
	near the location of the $v=0$ dark states after 7$\mu$s of evolution performed by the MDQT code. Plasma density is indicated in the legend, and $\delta=-\gamma$, $\delta_{D}=0$, $\Omega^{0}=\gamma$, and $\Omega^{0}_{D}=\gamma$. As the density increases beyond $10^{11}$\,m$^{-3}$, the dark-state populations become suppressed due to velocity changing collisions, which knock ions out of the velocity range for these dark states more quickly than dark-state coherences can develop.
	(\textbf{B}): Population in the ${^2P_{3/2}}$ state at $v=0$ after 7$\mu$s of evolution as a function of density.  The population saturates for $n\gtrsim 10^{13}$\,m$^{-3}$, indicating full collisionall suppression of the dark state.
	Data are the average of 99 runs, and the widths of the line/error bars represent the standard error of the results.}
	\label{fig:PPStateQTSimVaryDensity}
\end{figure*}

In order to test this, MDQT simulations were conducted  at a number of densities near this threshold with $\delta=-\gamma$, $\delta_{D}=0$, $\Omega^{0}=\gamma$, and $\Omega^{0}_{D}=\gamma$. The results for $P_{p}(v)$ after  $7\mu$s ($\sim 1000\gamma^{-1}$) of simulation are shown in Fig.~\ref{fig:PPStateQTSimVaryDensity}A. This time is long enough for the system to reach equilibrium. We clearly see suppression of this feature for $n\gtrsim 3\times 10^{11}$\,m$^{-3}$, as expected.  Figure~\ref{fig:PPStateQTSimVaryDensity}B shows $P_{p}(v=0)$ after  $7\,\mu$s as a function of density, which  saturates for $n\gtrsim 10^{13}$\,m$^{-3}$.  This is lower than typical UNP densities used in recent laser-cooling experiments ($10^{13}$\,m$^{-3}$ or greater) \cite{lgk2019}, implying that $v=0$ dark states had no impact on the cooling efficacy.

\begin{figure*}[t]
	\centering
	\includegraphics[scale=.6]{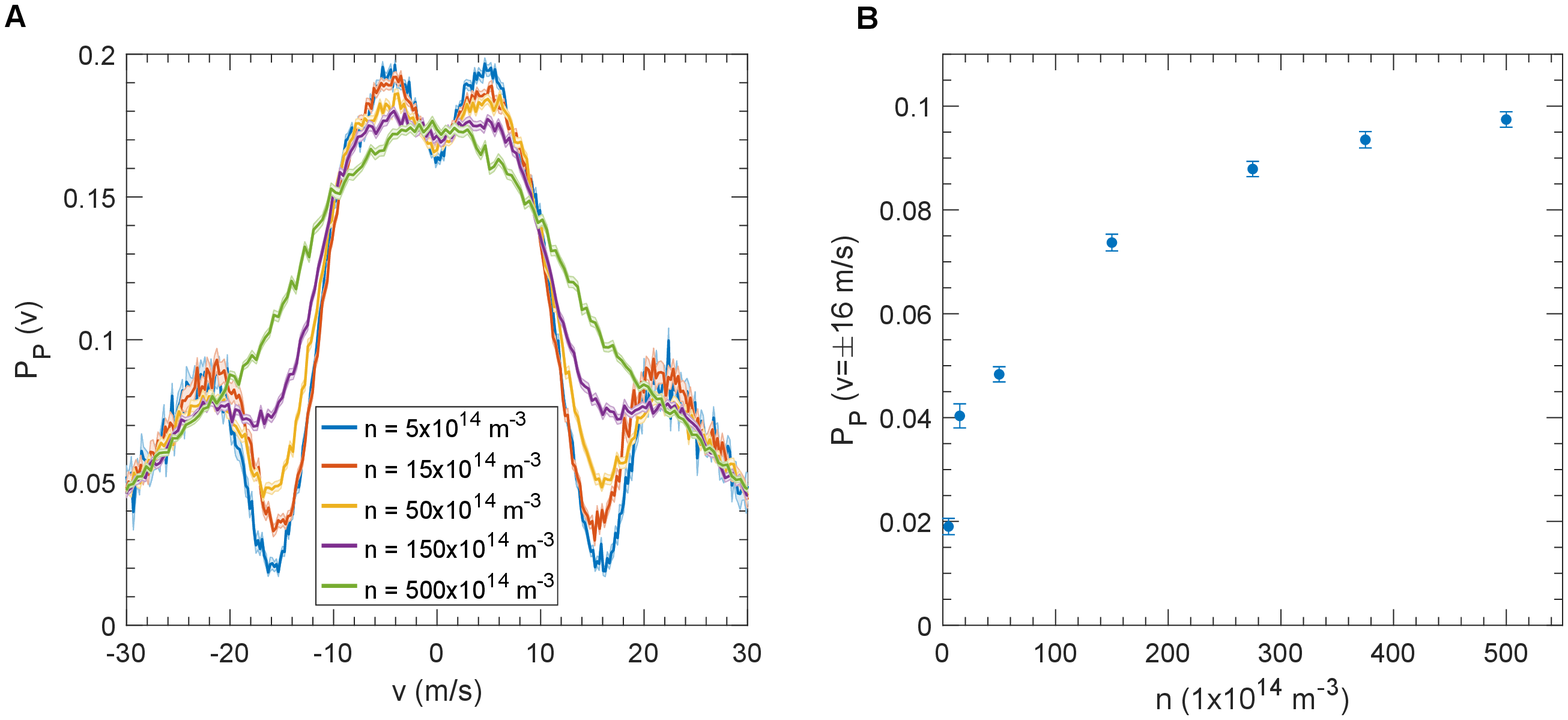}
	\caption[Collisional Supression of $v=\pm (\delta-\delta_{D})/(k-k_{D})$ Dark State.]{(\textbf{A}): Predictions from the MDQT code for population in the ${^2P_{3/2}}$ state versus $x$-velocity
 after 500\,ns of evolution. Density is indicated in the legend. As the density increases beyond $10^{15}$\,m$^{-3}$, the dark states at $v\sim\pm1.8\gamma/k=\pm16$\,m/s become increasingly collisionally suppressed.  (\textbf{B}): Population in the ${^2P_{3/2}}$ state at $v=\pm16$\,m/s after 500\,ns of evolution as a function of $n$.  The population saturates for $n\gtrsim2.5\times10^{16}$\,m$^{-3}$, indicating full collisional suppression of the dark states.
 Data are the average of 99 runs, and the widths of the line/error bars represent the standard error of the results.}
	\label{fig:PPStateV16QTSimVaryDensity}
\end{figure*}

On the other hand, the  dark states at $v\sim\pm1.8\gamma/k$ for $\delta_{D}=0$ (Fig.~\ref{fig:DarkStateTimeDevelopment}) develop on a timescale of $t\sim10\gamma^{-1}=70$\,ns and have a width $\delta v\sim3$\,m/s.  Estimating the suppression density in the same way as done for the $v=0$ dark state gives $n\ge 2\times 10^{16}$\,m$^{-3}$.  To test suppression of these states, MDQT simulations were performed for a range of densities between $5\times10^{14}$\,m$^{-3}$ and $5\times10^{16}$\,m$^{-3}$ for the same values of $\delta$, $\delta_{D}$, $\Omega^{0}$, and $\Omega^{0}_{D}$.  The resulting $P_{p}(v)$ curves after 500\,ns of plasma evolution are shown in Fig.~\ref{fig:PPStateV16QTSimVaryDensity}A.  The features are centered at $v=\pm16$\,m/s.  Plotting $P_{p}(v=\pm 16$\,m/s$)$ vs density ($n$) shows  that as $n$ increases, these features become increasingly suppressed as well, vanishing for $n\gtrsim2.5\times10^{16}$\,m$^{-3}$ (Fig.~\ref{fig:PPStateV16QTSimVaryDensity}B). This density is significantly higher than used in laser-cooling experiments \cite{lgk2019}, but for the chosen parameters, the velocity of these dark states is relatively high compared to mean ion thermal velocities ($v_{T}=\sqrt{k_{B}T/m_{i}}=7$\,m/s for $T=0.5$\,K). Thus, they do not prevent laser-cooling from being effective.

\section{Simulating Laser Cooling in a Uniform, Non-Expanding UNP}
\label{sec:SimResults}

\begin{figure*}[t]
	\centering
	\includegraphics[scale=.7]{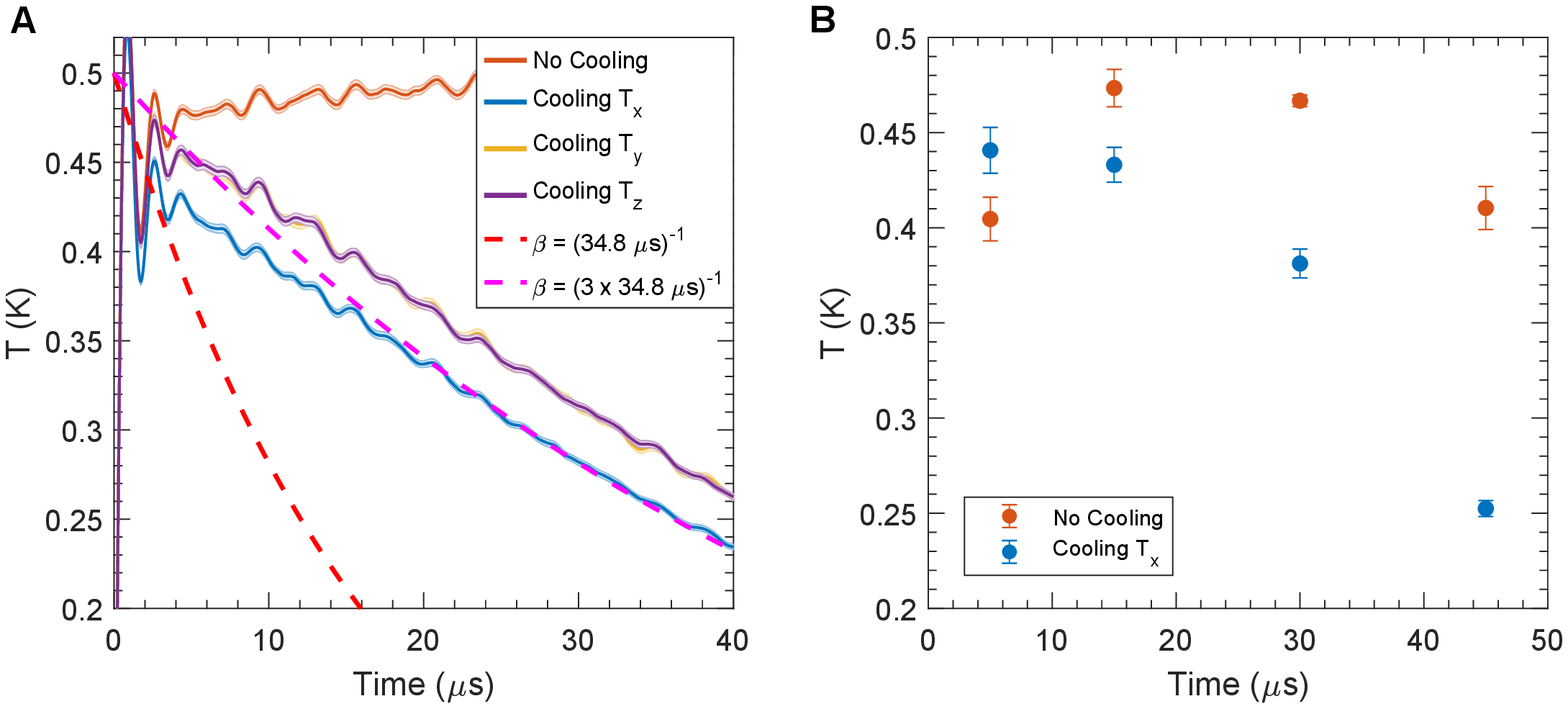}
	\caption[Temperature Along Cooled and Uncooled Axes in MDQT Simulation of Laser Cooling]{(\textbf{A}): Simulated ion temperatures, $T$\,(K), versus time for $n=2\times 10^{14}$\,m$^{-3}$ and $\kappa = 0.55$. For data with laser cooling, $\delta=-\gamma$, $\delta_{D}=\gamma$, $\Omega^{0}=\gamma$, and $\Omega^{0}_{D}=\gamma$. Even though the laser-cooling force is only applied along $x$, all three axes experience cooling due to collisional energy redistribution. Phenomenological curves for temperature decrease following $T(t)=T(0)e^{-2\beta t}$ for $T(0)=0.5$\,K and two different values of $\beta$ discussed in the text are also shown.
		Data are the average of 60 runs, and the widths of the line/error bars represent the standard error of the results.
 (\textbf{B}): Experimental measurements \cite{lgk2019} of the ion temperature along the laser-cooling axis, $T_x$\,(K), versus time for the center of a UNP with the same conditions as the above simulations.
}
	\label{fig:MDQTEnergiesSig0}
\end{figure*}


The MDQT simulation is well suited to describe laser cooling of ions in an UNP, however, there are important limitations. A simulation of a uniform density plasma with periodic spatial  boundary conditions, maps onto a plasma with no net hydrodynamic flow of particles and no overall plasma expansion with time. An experimentally realizable UNP, however, has a non-uniform density distribution and expands into surrounding vacuum \cite{kpr2007,lgk2019}. For a spherical Gaussian ion-density distribution, $n(r)=n(0)\mathrm{exp}(-r^2/2\sigma^2)$, where $r$ is the distance from the plasma center, the evolution of the plasma size and the hydrodynamic expansion velocity are given by \cite{kpr2007}
\begin{eqnarray}
\label{eq:plasmaexpansion}
  \sigma(t) &=& \frac{\sigma(0)}{\sqrt{1+t^2/\tau_{exp}^2}} \nonumber \\
  \vec{u}(\vec{r},t) &=&\vec{r} \frac{t/\tau_{exp}^2}{1+t^2/\tau_{exp}^2}.
\end{eqnarray}
Here, $t$ is the time after plasma creation and
$\tau_{exp}=\sqrt{m\sigma(0)^2/k_BT_e(0)}$ is a characteristic timescale for the expansion.
The simulation thus provides an accurate and valuable model of  conditions in the center of the plasma at early times $t<\tau_{exp}$, for which expansion velocity is small or vanishing.  Phenomena such as  adiabatic cooling and electron-ion energy exchange \cite{msk2015}, and the effects of expansion-induced Doppler shifts on laser cooling efficacy, are discussed in more detail in \cite{LanginPHD,lgk2019}.

An MDQT simulation of laser cooling was run with parameters matching recent UNP experiments \cite{lgk2019}, with $\Omega^{0}=\gamma$, $\Omega^{0}_{D}=\gamma$, $\delta_{D}=+\gamma$, and $\delta = -\gamma$. The density and screening parameter were  $n=2\times10^{14}$\,m$^{-3}$ and $\kappa=0.55$ respectively. This yields
$\omega_{pi}=2\times 10^{6}$\,s$^{-1}$ and $E_{c}/k_{B}=e^{2}/4\pi\epsilon_{0}a_{ws}k_{B}=1.6$\,K, which is
a characteristic Coulomb energy for two ions separated by the
Wigner-Seitz radius \cite{lsk2016}.

The natural timescale for ion motional dynamics for this plasma is
$\omega_{pi}^{-1}=0.5\,\mu$s.
The following quantities were recorded every time interval $\Delta
t=0.14\omega_{pi}^{-1}$,:\begin{itemize}
	\item The average ion kinetic energy along each axis, which is
	parameterized in terms of an effective temperature $T_{i,\nu}=m\langle
	v_{i,\nu}^{2}\rangle/2k_B$, where $\nu=x$, $y$, or $z$
	\item The total interaction energy from the shielded ion-ion potential.
	\item The velocity distribution along each axis
	$f(v_{x},v_{y},v_{z})$, with a bin spacing of $0.0043a_{ws}\omega_{pi}$.
	\item The $x$ velocity of each particle, along with its probability
	of being measured in the ${^2P_{3/2}}$ state, which allows calculation
	of $P_{p}(v)$.
\end{itemize}

Figure~\ref{fig:MDQTEnergiesSig0}A shows the simulated ion temperature along each
axis vs. time. At early times after
plasma creation ($t\le 5\,\mu$s), DIH and kinetic energy
oscillations are evident. These phenomena are characteristic of
equilibration of a plasma near or in the strongly coupled regime after a
rapid quench from non-interacting to interacting particles, which is a
good model of the photoionization, plasma-creation process
\cite{lsk2016,kpr2007,csk2004}.
The ions approach local thermal equilibrium at a temperature of
$T_i\approx 0.5$\,K. This equilibrium temperature is weakly dependent on the electron screening,
$\kappa$, but it  most strongly depends on ion
density, and is approximately
$T_i\approx E_{c}/3k_{B}$.

Without laser cooling, the temperature eventually stabilizes.
With laser cooling, the temperature decreases by a factor of two on a timescale of tens of
microseconds.  Only motion along the $x$-axis is directly laser
cooled, but the temperatures along the uncooled axes decrease at a rate
comparable to the cooled axis.
This is clear evidence of cross-axis, collisional thermalization.


The factor of 2 reduction in temperature observed in the simulation after 40\,$\mu$s of
cooling is large enough to measure with standard experimental probes of Sr UNPs
\cite{lgk2019,kpr2007,cgk2008}. Importantly, it occurs on a reasonably short
timescale compared to the timescale for expansion of the plasma for
experimentally realizable plasma parameters ($\tau_{exp}\sim 80$\,$\mu$s
\cite{lgk2019}). The simulation results are in good agreement with recent experimental
observations of laser cooling, as shown in Fig.~\ref{fig:MDQTEnergiesSig0}B. In the experimental data, the temperature without laser cooling decreases slightly at later times, which reflects the effects of plasma expansion and adiabatic cooling that are not included in the numerical simulation.

%

\subsubsection{Cooling and Thermalization Rates}
\label{subsubsec:MeasureThermRatesThroughLaserCooling}

Data from the simulation   can be fit by  rate equations in order to determine several phenomenological parameters describing various collision and laser-cooling processes. The rate equations,
which are equivalent to an approximate kinetic treatment \cite{ppr2004PRA,msk2015,lgk2019},
are
\begin{equation}
\begin{split}
\frac{\partial T_{\parallel}}{\partial t}&=-2\beta T_{x}+2\nu(T_{\perp}-T_{\parallel})-\frac{2}{3k_{B}}\frac{\partial U_{ii}}{\partial t}\\
\frac{\partial T_{\perp}}{\partial t}&=-\nu(T_{\perp}-T_{\parallel})-\frac{2}{3k_{B}}\frac{\partial U_{ii}}{\partial t}\\
\frac{\partial U_{ii}}{\partial t}&=-\mu\left[U_{ii}-U_{ii,Eq}\left(n,\bar{T},\kappa\right)\right]
\end{split}
\label{eq:tempEqsWithCorrHeat}
\end{equation}
The temperatures describing the velocity distributions parallel and perpendicular  to the cooling axis are $T_{\parallel}$ and $T_{\perp}$ respectively. $\beta$ characterizes the laser cooling force along the cooling axis for small velocity according to $F_x=-\beta m v$, which gives a temperature damping rate along that axis in the absence of any collisional effects of $T_x(t)=T_x(0)e^{-2\beta t}$.
The cross-axis thermalization rate is  $\nu$.

Equations \ref{eq:tempEqsWithCorrHeat} includes an energy source for the plasma that is important in and near the strongly coupled regime: the correlation energy, $U_{ii}<0$ \cite{ppr2004PRA,msk2015,lgk2019}, which is the potential energy compared to a system of the same density with no spatial correlations. In strongly coupled plasmas, spatial correlations exist that lower the potential energy. In a laser-cooled UNP experiment, as the plasma cools, the correlations increase, further decreasing the potential energy, and this is accounted for in the overall energy balance as an increase in thermal energy to balance the decrease in $U_{ii}$.  An MD simulation is necessary to calculate the time evolution of the correlation energy, but its influence on plasma temperature, averaged over a timescale long compared to $\omega_{pi}^{-1}$, can  be approximated by a model in which the correlation energy relaxes to its equilibrium,
$U_{ii,Eq}\left(n,T_i,\kappa\right)$, with a rate $\mu$ \cite{ppr2004PRA}. The equilibrium correlation energy, as a function of density, temperature, and screening parameter $\kappa$, can be taken from tabulated values \cite{hfd1997}. For determing $U_{ii,Eq}$, we approximate the ion temperature as the average $\bar{T}=\frac{T_{\parallel}+2T_{\perp}}{3}$.

\begin{figure}[!h]
	\centering
	\includegraphics[scale=.6]{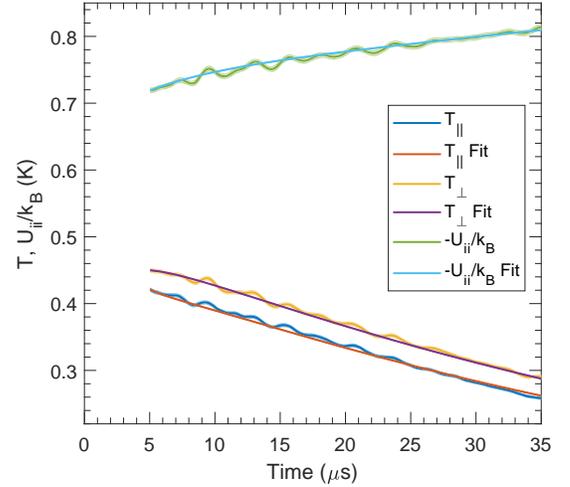}
	\caption[Fits of $T_{\parallel}$, $T_{\perp}$, and $-T_{corr}$ to Thermalization Model.]{Results of fitting MDQT data (Fig.~\ref{fig:MDQTEnergiesSig0}) to Eq.~\ref{eq:tempEqsWithCorrHeat} with $\beta$, $\nu$, and $\mu$ as free parameters. Data are the average of 60 runs, and the widths of the line/error bars represent the standard error of the results. }
	\label{fig:thermRates}
\end{figure}

The rate equations (Eqs.\ \ref{eq:tempEqsWithCorrHeat}) do not factor in any of the oscillatory behavior observed at early times during DIH. Thus, we fit the simulation data for $t>5$\,$\mu$s. Results are shown in Fig.~\ref{fig:thermRates}, and the fit parameter values are\begin{itemize}
	\item $\beta=(3.30\pm0.01)\times10^{4}$\,s$^{-1}$
	\item $\nu=(0.116\pm0.002)\,\omega_{pi}$
	\item $\mu=(0.088\pm0.002)\,\omega_{pi}$
\end{itemize}

As a check, we determine the temperature relaxation rate for the same cooling parameters
in the absence of particle interactions with a pure QT simulation of laser cooling (data not shown). This yields a decay of the temperature along the cooling axis following $T_x(t)=T_x(0)e^{-2\beta t}$, with $\beta=(34.8\,\mu \mathrm{s})^{-1}=2.9\times 10^4\, \mathrm{s}^{-1}$, which is close to the value determined from the full MDQT simulation.
This is also close to an estimation of $\beta$ based on a measurement of the optical pumping rate to $^2D_J$ states during laser cooling \cite{lgk2019} (supplementary material).

Note from Fig.\ \ref{fig:MDQTEnergiesSig0} that the temperatures actually fall at approximately one third the rate for a one-dimensional system in the absence of particle interactions:
$T(t)=T(0)e^{-2\beta t/3}$, rather than $T(t)=T(0)e^{-2\beta t}$. This reflects the fact that laser-cooling only acts on one degree of freedom while collisions rapidly redistribute energy between perpendicular and parallel dimensions.

The perpendicular temperature lags behind the parallel temperature during cooling. This makes it possible to use the temperature curves to determine the cross-thermalization rate $\nu$. Because UNP ions are an excellent realization of the YOCP model used for describing high-density strongly coupled plasmas,  experimental and numerical determination of $\nu$ in this system is of significant interest. The fitted value for $\nu$ agrees well with direct MD simulations from \cite{bda2017}, which found $\nu \approx 0.1\,\omega_{pi}$. A thorough experimental and numerical study of cross-thermalization rates in laser-cooled UNPs is the subject of future work.

\begin{figure}[!h]
	\centering
	\includegraphics[scale=.58]{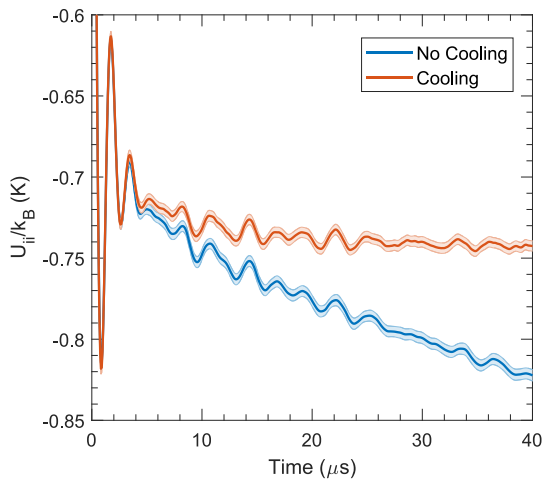}
	\caption[Correlation Temperature With and Without Laser Cooling]{Correlation energy with and without laser cooling for conditions used in Fig.~\ref{fig:MDQTEnergiesSig0}.  When laser cooled, the ions become more spatially correlated, and the potental energy decreases faster than when there is no cooling.  Due to energy conservation, this effect introduces an additional heating term to the differential equations determining the temperature evolution in a laser-cooled plasma  (Eqs.~\ref{eq:tempEqsWithCorrHeat}).
	Data are the average of 60 runs, and the widths of the line/error bars represent the standard error of the results.}
	\label{fig:UPotVsTimeCoolAndNoCool}
\end{figure}

We can also use simulation results to investigate the behavior of the correlation energy during laser cooling.  $U_{ii}(t)$ is easily  determined from the recorded positions of the plasma ions by calculating the difference between the potential energy at time $t$ and the initial potential energy (for which there were no spatial correlations). $U_{ii}(t)$ determined in this fashion with and without laser cooling is plotted in Fig.~\ref{fig:UPotVsTimeCoolAndNoCool}.  Without laser cooling, $U_{ii}(t)$ remains roughly constant after the initial DIH equilibration phase. Decreasing temperature resulting from laser cooling, however, increases the spatial correlations and lowers the potential energy.

\section{Conclusion}
\label{sec:SimSummary}
We have developed a combined MDQT code for simulating laser-driven processes in a collisional system. We have applied this code to investigate the effect of one-dimensional laser cooling of the ions within an UNP. The MDQT simulations demonstrate that  laser-cooling  can reduce the temperature by a factor of two along all axes in $\sim 40\,\mu$s, in agreement with recent experimental results \cite{lgk2019}. The simulation confirms that collisions  isotropize energy across all degrees of freedom efficiently on this laser-cooling timescale. We also observe that collisions suppress the development of dark states, which might otherwise inhibit laser-cooling.

More generally, this code can be adapted to describe any many-body system in which laser manipulation
of internal quantum states and velocity-changing collisions occur on similar timescales.
In UNPs, there are other important processes that can be studied with this tool, such
as the laser-induced fluorescence probe used for thermometry \cite{cgk2008} and  the development and relaxation of spin-velocity correlations used for measuring collision rates, diffusion\textbf{,} and velocity auto-correlation functions \cite{bcm2012}.



%
%

While this simulation is a powerful tool for the reasons described above, it is in some sense incomplete, as it cannot realistically describe an inhomogeneous system and does not  account for the expansion of the plasma.  For example, adiabatic cooling and  density reduction associated with the expansion are noticeably absent.
The effects of laser-cooling forces on the expansion \cite{lgk2019}, likewise, cannot be investigated with this tool.
A full simulation of laser cooling an UNP is an exciting scientific challenge that would require a multiscale approach in which the MDQT simulation acts at the lowest level while either a hydrodynamic or kinetic code handles the macroscopic expansion.

\section*{Acknowledgments}
\label{sec:Acknowledgments}
This work was supported by the Air Force Office of Scientific Research through grant FA9550-17-1-0391 and the National Science Foundation Graduate Research Fellowship Program under Grant No. 1842494. Any opinions, findings, and conclusions or recommendations expressed in this material are those of the authors and do not necessarily reflect the views of the National Science Foundation.
One of the authors (DV) is also grateful for the support received from the National Science Foundation through grants PHY-1831977 and HRD-1829184.

%

\appendix

\section{QT Details}
\label{app:QTdetails}
\subsection{Effective Hamiltonian}
For modeling one-dimensional laser cooling of Sr ions,
the effective Hamiltonian (Eq.\ (\ref{eq:psiEv})) is

\begin{widetext}
\begin{equation}
\label{eq:HqteffExample}
\begin{split}
H^{QT}_{eff}&=\hbar\omega(|3\rangle\langle 3|+|4\rangle\langle 4|+|5\rangle\langle 5|+|6\rangle\langle 6|)+\hbar\omega_{D}(|7\rangle\langle 7|+|8\rangle\langle 8|+|9\rangle\langle 9|+|10\rangle\langle 10|+|11\rangle\langle 11|+|12\rangle\langle 12|)\\
&-\frac{\hbar}{2}\left(|2\rangle\langle3|+\frac{|1\rangle\langle 4|}{\sqrt{3}}+h.c.\right)\left(\Omega\exp\left[-i(\nu+kv)t\right]+\Omega^{*}\exp\left[i(\nu+kv)t\right]\right)\\
&-\frac{\hbar}{2}\left(\frac{|2\rangle\langle 5|}{\sqrt{3}}+|1\rangle\langle 6|+h.c.\right)\left(\Omega\exp\left[-i(\nu-kv)t\right]+\Omega^{*}\exp\left[i(\nu-kv)t\right]\right)\\
&-\frac{\hbar}{2}\left(\frac{|10\rangle\langle 3|}{\sqrt{15}}+\frac{|9\rangle\langle 4|}{\sqrt{5}}+\frac{\sqrt{2}|8\rangle\langle 5|}{\sqrt{5}}+\frac{\sqrt{2}|7\rangle\langle 6|}{\sqrt{3}}+h.c.\right)
\left(\Omega_{D}\exp\left[-i(\nu_{D}-k_{D}v)t\right]+\Omega_{D}^{*}\exp\left[i(\nu_{D}-k_{D}v)t\right]\right)\\
&-\frac{\hbar}{2}\left(\frac{|9\rangle\langle 6|}{\sqrt{15}}+\frac{|10\rangle\langle 5|}{\sqrt{5}}+\frac{\sqrt{2}|11\rangle\langle 4|}{\sqrt{5}}+\frac{\sqrt{2}|12\rangle\langle 3|}{\sqrt{3}}+h.c.\right)\left(\Omega_{D}\exp\left[-i(\nu_{D}+k_{D}v)t\right]+\Omega_{D}^{*}\exp\left[i(\nu_{D}+k_{D}v)t\right]\right)\\
& -i\sum_{k=1}^{18}\frac{\gamma_{k}}{2}c_{k}^{\dagger}c_{k}
\end{split}
\end{equation}
\end{widetext}
\noindent where $v=p_x/m$, $\hbar\omega$ is the energy of the ${^2P_{3/2}}$ state, $\hbar\omega_{D}$ is the energy of the ${^2D_{5/2}}$ state, $\Omega^{0}$ is the laser-induced Rabi frequency between states $S$ and $P$ for a hypothetical transition with Clebsch-Gordon (C-G) coefficient of 1. $\Omega^{0}_{D}$ is the same but for coupling between $D$ and $P$. $\nu$ and $\nu_{D}$ refer to the frequency of the coupling lasers. $\gamma_{k}$ and $c_{k}$ refer to the 18 decay paths indicated in Fig.~\ref{fig:FullSrLevelDiagram}D and Fig.~\ref{fig:FullSrLevelDiagram}E. We have included the relevant C-G coefficients and Doppler shifts of magnitude $kv$ and $k_{D}v$ where $k$ is the wavenumber for the ${^2S_{1/2}}\rightarrow{^2P_{3/2}}$ transition and $k_{D}$ is the wavenumber for the ${^2D_{5/2}}\rightarrow{^2P_{3/2}}$ transition.
We incorporate the spatial dependence of the light fields as
 $\Omega=\Omega^0\exp[-ikx]$ and $\Omega^{*}=\Omega^0\exp[ikx]$.
The  signs in the exponents are consistent with the $\sigma^{+}$ wave for the ${^2S_{1/2}}\rightarrow{^2P_{3/2}}$ transition  propagating from positive $x$ to negative $x$.
Similarly, $\Omega_{D}=\Omega^0_{D}\exp[-ik_Dx]$ and $\Omega^{*}_{D}=\Omega^0_{D}\exp[ik_Dx]$. Figure \ref{fig:FullSrLevelDiagram}B provides the state labels, while Figs. \ref{fig:FullSrLevelDiagram}D and \ref{fig:FullSrLevelDiagram}E indicate the 18 decay paths and decay rates described by the last term in Eq. (\ref{eq:HqteffExample}).

To eliminate the time dependence, it is customary to transform to a basis set where wavefunctions are rotating with the light field (including the Doppler shift) and neglect resulting terms proportional to $\sim\exp\left[2i\nu t\right]$, since we are not interested in dynamics on the timescale $\nu^{-1}$.  However, in this case we cannot completely eliminate the time dependence.  This is because the $m_{j}=\pm 1/2$ states in the ${^2D_{5/2}}$ manifold are coupled to the $m_{j}=\pm 1/2$ states in the ${^2S_{1/2}}$ manifold through two different ${^2P_{3/2}}$ states.  For example, states $|1\rangle$ and $|9\rangle$ are coupled through both $|6\rangle$ and $|4\rangle$, meaning that there is some ambiguity regarding which rotating field to use for the transformation of these states.  In this case, we choose to transfer $|9\rangle$ and $|10\rangle$ to the frame rotating with the $\sigma^{+}$ ${^2D_{5/2}}\rightarrow{^2P_{3/2}}$ laser.  The resulting Hamiltonian after the unitary transformation to the rotating frame is

\begin{widetext}
\begin{equation}
\begin{split}
\frac{H^{QT}_{eff}}{\hbar}&=(-\delta-vk)(|3\rangle\langle3|+|4\rangle\langle4|)+(-\delta+vk)(|5\rangle\langle5|+|6\rangle\langle6|)+(-\delta+\delta_{D}+(k-k_{D})v)(|7\rangle\langle7|+|8\rangle\langle8|)\\
&+(-\delta+\delta_{D}+(-k+k_{D})v)(|11\rangle\langle11|+|12\rangle\langle12|)+(-\delta+\delta_{D}+(-k-k_{D})v)(|9\rangle\langle9|+|10\rangle\langle10|)\\
&+\left(\frac{\Omega^{*}}{2}|2\rangle\langle3|+\frac{\Omega^{*}}{2\sqrt{3}}|1\rangle\langle4|+\frac{\Omega^{*}}{2}|1\rangle\langle6|+\frac{\Omega^{*}}{2\sqrt{3}}|2\rangle\langle 5|+h.c\right)\\
&+\left(\frac{\sqrt{2}\Omega_{D}^{*}}{2\sqrt{3}}|7\rangle\langle6|+\frac{\sqrt{2}\Omega_{D}^{*}}{2\sqrt{5}}|8\rangle\langle5|+\frac{\Omega_{D}^{*}}{2\sqrt{5}}|9\rangle\langle4|+\frac{\Omega_{D}^{*}}{2\sqrt{15}}|10\rangle\langle 3|+h.c\right)\\
&+\left[\exp\left[2i(k+k_{D})vt\right]\left(\frac{\Omega_{D}^{*}}{2\sqrt{15}}|9\rangle\langle 6|+\frac{\Omega_{D}^{*}}{2\sqrt{5}}|10\rangle\langle 5|\right)+h.c\right]\\
&+\left(\frac{\sqrt{2}\Omega_{D}^{*}}{2\sqrt{5}}|11\rangle\langle 4|+\frac{\sqrt{2}\Omega_{D}^{*}}{2\sqrt{3}}|12\rangle\langle 3|+h.c\right) -i\sum_{k=1}^{18}\frac{\gamma_{k}}{2}c_{k}^{\dagger}c_{k}
\end{split}
\label{eq:Heff}
\end{equation}
\end{widetext}

\noindent where the remaining time dependence results from the difference in frequency between the chosen rotating frame and the frame rotating with the `alternate' paths coupling the ${^2S_{1/2}}$ and ${^2D_{5/2}}$ states.

To determine the optical force for the one-dimensional laser cooling configuration,  consider the application of Eq.~\ref{eq:ForceProfileFromTrace} to the $|2\rangle\langle 3|$ and $|3\rangle\langle 2|$ term of $H^{QT}_{eff}$, which yields
\begin{eqnarray}
  \langle F^{QT}_{x,23}\rangle &=& -\left\langle\frac{\left[p_x,H_{QT,23}\right]}{i\hbar}\right\rangle \\
  &=& \langle\psi|\left[\frac{\partial}{\partial x}\left(\frac{\hbar\Omega^{*}}{2}|2\rangle\langle3|+\frac{\hbar\Omega}{2}|3\rangle\langle2|\right)\right]|\psi\rangle  \nonumber
\end{eqnarray}
Inserting $\Omega=\Omega^{0}\exp[-ikx]$ (the minus sign is due to the fact that the $\sigma^{+}$ wave for the S$\rightarrow$P transition is propagating to the left) and $\Omega^{*}=\Omega^{0}\exp[ikx]$, we get:
\begin{eqnarray}
 \langle F^{QT}_{x,23}\rangle &=&  \frac{ik\hbar\Omega^{0}}{2}\left(\langle\psi|2\rangle\langle 3|\psi\rangle-\langle\psi|3\rangle\langle2|\psi\rangle\right)\nonumber \\
      &=& -k\hbar\Omega^{0} Im[\rho_{32}]
\end{eqnarray}

After considering all such terms in the Hamiltonian, the total force is written as
\begin{equation}
\begin{split}
\label{eq:Forcei}
\langle F^{QT}_{x}\rangle &= k\hbar\Omega^0\left(-Im[\rho_{32}]+Im[\rho_{61}]\right)\\
&
+\frac{k\hbar\Omega^0}{\sqrt{3}}\left(-Im[\rho_{41}]+Im[\rho_{52}]\right)\\
& +\frac{\sqrt{2}k_{D}\hbar\Omega^0_{D}}{\sqrt{3}}\left(Im[\rho_{67}]-Im[\rho_{3\,12}]\right)\\
&
+\frac{\sqrt{2}k_{D}\hbar\Omega^0_{D}}{\sqrt{5}}\left(Im[\rho_{58}]-Im[\rho_{4\,11}]\right)\\
&
+\frac{k_{D}\hbar\Omega^0_{D}}{\sqrt{5}}\left(Im[\rho_{49}]-Im[\rho_{5\,10}]\right)\\
&
+\frac{k_{D}\hbar\Omega^0_{D}}{\sqrt{15}}\left(Im[\rho_{3\,10}]-Im[\rho_{69}]\right)
\end{split}
\end{equation}
where we have suppressed the ion index $i$ on all quantities.

In a time step $\Delta t_{QT}$ during which a quantum jump has \emph{not} occurred, the momentum  changes by
\begin{equation}
\label{eq:dpx}
 \displaystyle \Delta \vec{p}=\langle\vec{F}^{QT}\rangle\,\Delta t_{QT}+\frac{\Delta \vec{p}^{MD}}{N}.
\end{equation}
 $\langle \vec{F}^{QT}\rangle=\langle F^{QT}_{x}\rangle\,\hat{x}$ is the optical force on the particle, which only has an x-component given by Eq.\ \ref{eq:Forcei}. \textbf{$\Delta \vec{p}^{\,MD}$} is the most recent calculation of the classical-force momentum kick passed to the QT algorithm. In a time step during which a quantum jump has occurred, the momentum changes by
 \begin{equation}
 \label{dpx,elec}
 \displaystyle \Delta \vec{p}=\Delta \vec{p}_{recoil}+\frac{\Delta \vec{p}^{\,MD}}{N}
 \end{equation}
where $\Delta \vec{p}_{recoil}$ is the appropriate photon recoil for the  photon emitted during the quantum jump.



\subsection{Execution of the Quantum Trajectories Algorithm}
\label{subsec:executeQTalgorithm}
The quantum trajectories algorithm for evolving both the momentum and the wavefunction  for a single particle $i$ is executed as follows.
Given a wavefunction $|\psi_i(t)\rangle$ and momentum $\vec{p}_{i}(t)$ we obtain $|\psi_{i}(t+\Delta t_{QT})\rangle$ and $\vec{p}_{i}(t+\Delta t_{QT})$ in the following way:\begin{itemize}
	\item (1) Pick a random number $r$ between 0 and 1.
	\item (2) Calculate $\Delta P$ using Eq.~\ref{eq:dp}.  If $\Delta P<r$, there is no jump, move to Step (3a).  If not, then there is a jump, move to Step (3b).
	\item (3a) Calculate $\Delta\vec{p}_{i}$ using Eq.~\ref{eq:dpx}.
	Set $\displaystyle\vec{p}_{i}(t+\Delta t_{QT})=\vec{p}_{i}(t)+\Delta \vec{p}_{i}$
	\item (4a) Using $\vec{p}_{i}(t)$, 
	calculate $H^{QT}_{eff}$ using Eq.~\ref{eq:Heff}.
	\item (5a) Use $H^{QT}_{eff}$ and Eq.~\ref{eq:psiEv} to determine $|\psi_i(t+\Delta t_{QT})\rangle$ and $\rho_i(t+\Delta t_{QT})=|\psi_i(t+\Delta t_{QT})\rangle\langle\psi_i(t+\Delta t_{QT})|$ with a $4^{th}$ order Runge-Kutta algorithm.
	\item (6a) Go back to Step 1
	\item (3b) Pick a random number $r_{2}$.  If $\Delta P_{k=1}<r_{2}$, the $k=1$ transition indicated in Fig.~\ref{fig:FullSrLevelDiagram}D occurs and the particle state jumps to $|2\rangle=|\psi_i(t+\Delta  t_{QT})\rangle$.  Else if $\Delta P_{k=2}+\Delta P_{k=1}<r_{2}$, the $k=2$ transition occurs, and so on.  For example, if the $k=10$ jump is selected, the transition is to the $|8\rangle=|\psi_i(t+\Delta t_{QT})\rangle$ state. (See Figs. \ref{fig:FullSrLevelDiagram} (D) and (E) for transition labels.)
	\item (4b) Randomly decide the direction of the recoil kick.
	\item (5b) If the state after the jump is either $|2\rangle$ or $|1\rangle$, then set $\vec{p}_{i}(t+\Delta t_{QT})=\vec{p}_{i}(t)+\frac{\Delta \vec{p}_i^{\,MD}}{N}\pm \hbar k\hat{x}$.  Else, set $\vec{p}_{i}(t+\Delta t_{QT})=\vec{p}_{i}(t)+\frac{\Delta \vec{p}_i^{\,MD}}{N}\pm \hbar k_{D}k\hat{x}$. (See Eq.\,\ref{dpx,elec}.)
	\item (6b) Go back to Step 1.
\end{itemize}

Here we list numerical values used in the Sr$^{+}$ laser-coupling simulation:

\begin{itemize}
	\item $\lambda = 407.8865$\,nm
	\item $k = 2\pi/\lambda=1.54\times10^{5}$cm$^{-1}$
	\item $\lambda_{d} = 1033.0139$\,nm
	\item $k_{d}=6.0825\times10^{4}$cm$^{-1}$
	\item $\gamma=1.41\times10^{8}$s$^{-1}=(7.09$\,ns$)^{-1}$
	\item $\gamma_{D} = 8.7\times 10^{6}$\,s$^{-1}$
\end{itemize}



\end{document}